\documentclass[preprint,showpacs,preprintnumbers,amsmath,amssymb,showkeys,aps]{revtex4}
\usepackage{graphicx}
\usepackage{dcolumn}
\usepackage{bm}
\usepackage[ansinew]{inputenc}
\usepackage[T1]{fontenc}
\newcommand{\A}{{\bf A}~}
\newcommand{\B}{{\bf B}~}

\begin{document}

\preprint{Anticrossing}
\title{Direct Observation of Controlled Coupling in an Individual Quantum Dot Molecule}

\author{H. J. Krenner}
\author{M. Sabathil}
\author{E. C. Clark}
\author{A. Kress}
\author{D. Schuh}
\author{M. Bichler}
\author{G. Abstreiter}
\author{J. J. Finley}

\affiliation{Walter Schottky Institut and Physik Department, Technische Universit\"at M\"unchen, Am Coulombwall 3, D-85748 Garching, Germany}

\date{\today}

\begin{abstract}

We report the direct observation of quantum coupling in individual quantum dot molecules and its manipulation using static electric fields.  A pronounced anti-crossing of different excitonic transitions is observed as the electric field is tuned.  Comparison of our experimental results with theory shows that the observed anti-crossing occurs between excitons with predominant spatially \emph{direct} and \emph{indirect} character.  The electron component of the exciton wavefunction is shown to have molecular character at the anti-crossing and the quantum coupling strength is deduced optically.  In addition, we determine the dependence of the coupling strength on the inter-dot separation and identify a field driven transition of the nature of the molecular ground state.

\end{abstract}

\pacs{78.67.Hc  71.35.Cc  73.21.La  78.55.Cr}

\maketitle
The realization of robust and scalable hardware for quantum information processing is one of the most challenging goals in modern solid-state physics. Single charge excitations (excitons) in semiconductor quantum dots represent a particularly attractive quantum bit (qubit) since they can be coherently manipulated using ultrafast laser pulses over timescales much shorter than their decoherence time.\cite{Biolatti,SFB,
Evi}  However, for an isolated dot such approaches are fundamentally limited to one or two qubit operations\cite{Li} with little or no prospects for further scalability. As a result, attention has recently shifted towards few dot nanostructures that have potential for building more complex quantum processors.\cite{Biolatti,Villas,Sche} One of the most promising system for implementing two qubits is a "quantum dot molecule" (QDM) consisting of a vertically stacked pair of GaInAs islands formed via strain-driven self-assembly in a GaAs matrix.\cite{Xie} Recent experiments have revealed radiatively limited dephasing for excitons in QDMs\cite{BorriQDM}, supporting their potential to perform basic quantum operations using an intrinsically scalable architecture. However, until now little tangible proof for quantum coupling in QDMs has appeared, despite first indications of tunnel coupling from single dot photoluminescence (PL) investigations.\cite{Bay,Ort,OrtPhE2} The findings of these studies are inconsistent with recent theoretical modeling\cite{Bester}, which underscores the need for realistic calculations that include strain, quantum and Coulomb couplings. Moreover, the lack of any direct experimental observations or consensus of opinion calls for investigation of \textit{tunable} systems for which the inter-dot coupling can be manipulated.
Electric field perturbations have been suggested as a tuning mechanism of the exciton spectrum via the Quantum Confined Stark Effect (QCSE) providing a useful method to control the strength of qubit-qubit coupling.\cite{Biolatti,Villas,Sht,OrtPhE2,Shen}

In this letter we report the direct spectroscopic observation of the quantum coupling in individual QDMs and its manipulation using static electric fields.  A clear anti-crossing of two exciton states is observed as the magnitude of the electric field ($F$) along the QDM axis is tuned.  These observations are repeated for eight single QDMs, providing a clear and general fingerprint of such controlled quantum coupling.  Furthermore, by comparing our experimental results with realistic calculations of the excitonic states we show that the observed anti-crossing occurs between excitons that have predominantly \emph{direct} and \emph{indirect} character with respect to the spatial distribution of the electron and hole wavefunctions.  Good quantitative agreement is obtained between experiment and theory demonstrating a modification of the character of the QDM ground state from direct to indirect as electric field increases.

\begin{figure}[tbhp]
    \begin{center}
       \includegraphics[width=0.8\columnwidth]{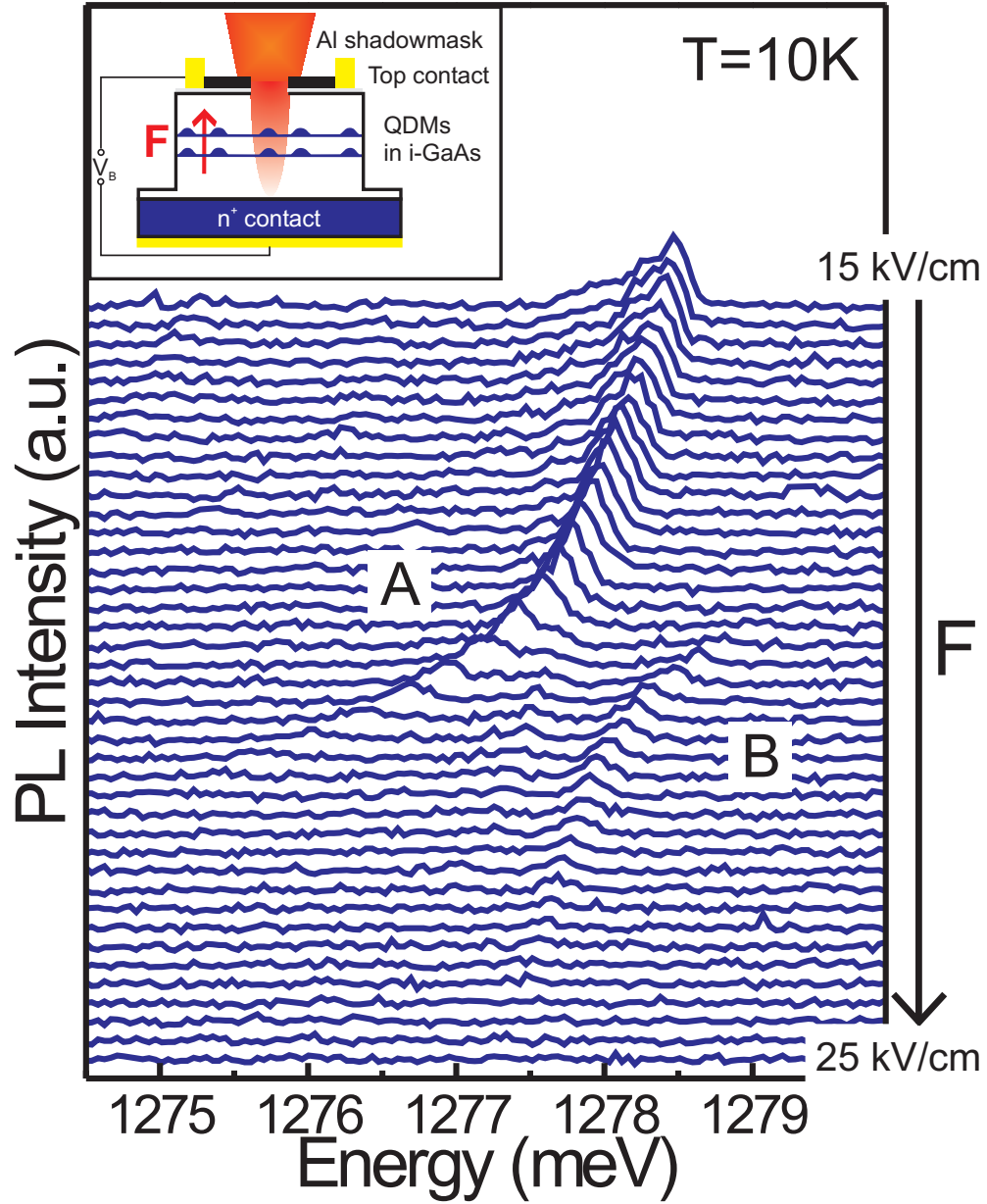}
   \end{center}
   \caption{\label{fig1}(Color online) {\it Main panel}: Photoluminescence of a single QDM as a function of the applied electric field ranging from 15 to 25$\mathrm{{kV}/{cm}}$. The two excitons, \A and \textbf{B}, show a clear anti-crossing with a minimum splitting of $\Delta E$=1.4meV at $\sim 18.8\mathrm{{kV}/{cm}}$. {\it Inset}: Schematic of the investigated \textit{n-i} Schottky diodes. The QDMs are embedded in the intrinsic GaAs layer with the electric field ($F$) oriented along the stack axis, parallel to the growth direction ([001]).}
\end{figure}

To investigate the influence of electric field perturbation on a QDM we performed photoluminescence (PL) spectroscopy on single pairs of self-assembled QDMs embedded in {\it n-i~}Schottky junctions. Samples were grown by molecular beam epitaxy on $n^+-$doped [001] GaAs substrates.  The epitaxial layer sequence was as follows:  firstly, a highly doped $n^+$-back contact was grown followed by 110nm of undoped GaAs. The two layers of GaInAs dots were then deposited separated by a nominally $d=10\mathrm{nm}$ thick GaAs spacer.  The two dot layers consisted of 8ML (lower dot layer) and 7ML (upper dot layer) of Ga$_{0.5}$In$_{0.5}$As deposited at 530$^\mathrm{o}$C.
In a first growth step, a single layer of islands is deposited and capped by a GaAs spacer. These buried islands then act as stressors, providing preferential nucleation sites for dots in the second layer, directly above the underlying dot, to create a QDM.\cite{Xie} Finally, the samples were completed with $120\mathrm{nm}$ of undoped GaAs.  Cross sectional transmission electron microscopy (TEM) measurements revealed lens shaped dots with vertical and lateral sizes of $\sim 4\mathrm{nm}$ and $\sim 20\mathrm{nm}$ respectively, the upper dot tending to be slightly larger compared with the lower. The samples were processed into photo diodes equipped with opaque aluminium shadow masks patterned with $\sim 500\mathrm{nm}$ diameter apertures to isolate single QDMs for optical investigation. A schematic of the device structure is given in the inset of Fig. \ref{fig1}. By applying a bias voltage between the $n$-contact and the Schottky-gate the axial electric field ($F$) can be tuned from 0 to $\sim250\mathrm{kV/cm}$. For low electric fields ($F<30\mathrm {kV/cm}$) the exciton radiative lifetime is shorter than the carrier tunneling escape time out of the QDMs and PL-experiments could be performed.

\begin{figure}[htbp]
    \begin{center}
        \includegraphics[width=0.8\columnwidth]{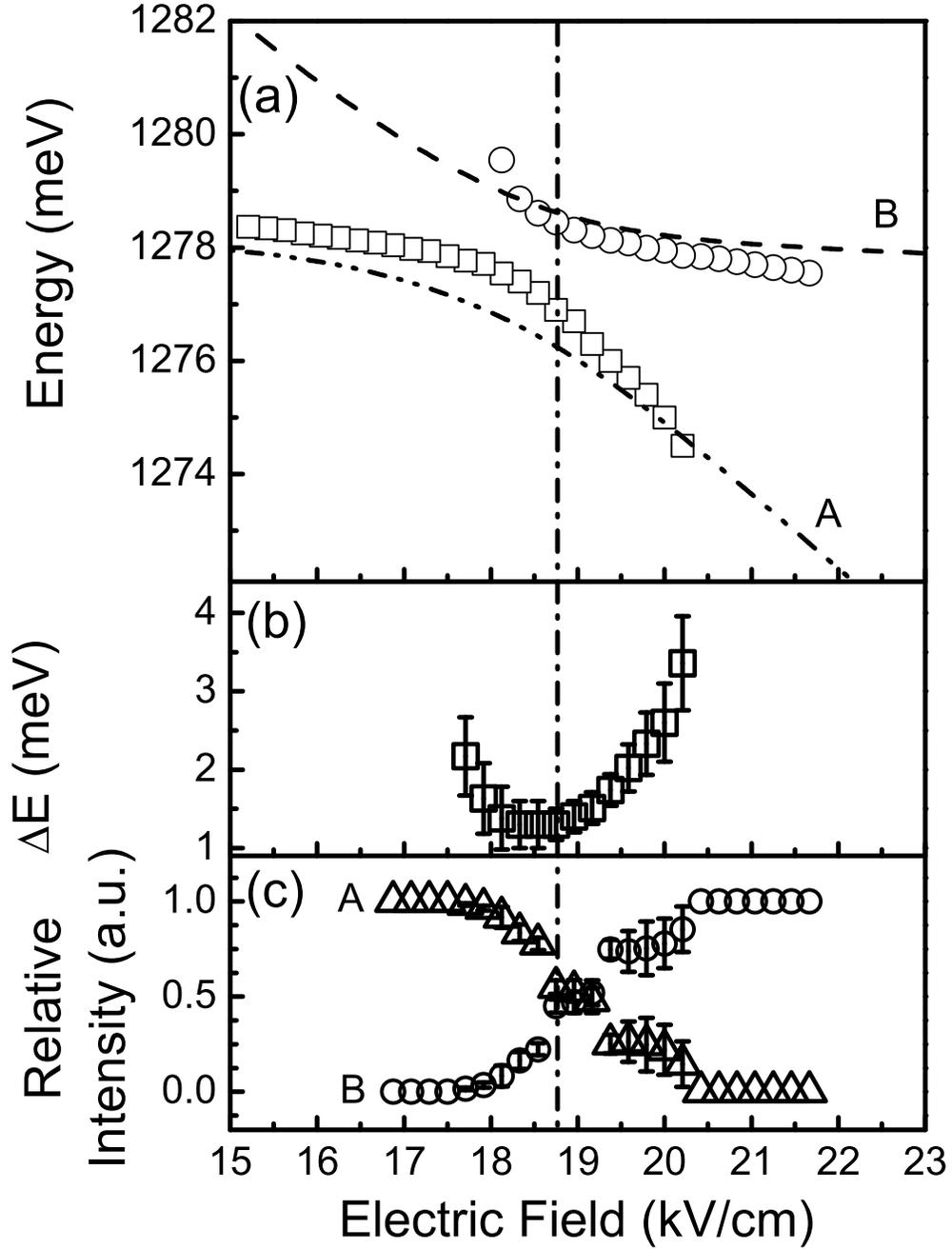}
    \end{center}
    \caption{\label{fig2}(a) Calculated and experimentally obtained exciton energies of state \A and \B as a function of the electric field for model QDMs as described in the text.(b) Energy splitting between \A and \B and (c) Relative intensities $I_{\A/\B}/(I{_\A}+I{_\B})$. The vertical line indicates the field-coincidence of minimum $\Delta E$ and equal relative intensities, illustrating the clear anti-crossing.}
\end{figure}

Typical PL-spectra obtained  at low temperatures (T$=10$K) from an individual QDM as a function of axial electric field are presented in the main panel of Fig. \ref{fig1}.  The spectra were recorded with very low excitation power ($P_{ex}\sim$2.5Wcm$^{-2}$) to ensure that only single exciton species are generated.\cite{countrate}  At low electric fields ($F\leq 17\mathrm{kV/cm}$) a dominant spectral line, labeled \A in Fig. \ref{fig1}, is observed.  This feature shifts weakly with increasing $F$ due to the QCSE until, at an electric field of $F \sim 18.5\mathrm{kV/cm}$, the shift rate of \A strongly increases and its intensity quenches rapidly.  Over the same electric field range a second peak, labeled \B in Fig. \ref{fig1}, exhibits precisely the opposite behavior: it is weak and shifts rapidly for $F\lesssim 18.5\mathrm{kV/cm}$ whereafter the shift rate decreases and the peak gains intensity relative to \textbf{A}.  The relative intensities of \A and \B are found to be anti-correlated as presented in Fig. \ref{fig2}(c) which shows $I_{\A/\B}/(I{_\A}+I{_\B})$.  For $F<18.5\mathrm{kV/cm}$ peak \A has greater oscillator strength, at $F=18.8\mathrm{kV/cm}$ both branches are equally intense and at high fields peak \B dominates.  The overall reduction in the absolute PL intensity arises from field induced suppression of carrier capture and the transition of the groundstate to an optically inactive state. The peak positions and energy splitting of peaks \A and \B are presented in Figs. \ref{fig2}(a) and (b) as a function of $F$.  Clearly, the minimum energy splitting ($\Delta E = 1.4\mathrm{meV}$) occurs at $F=18.8\mathrm{kV/cm}$. When considered together the data presented in Figs. \ref{fig1} and \ref{fig2} demonstrate that \A and \B anti-cross as they are tuned into resonance. Such anti-crossings are a clear signature of a tunable, coupled quantum system. Similar results were obtained from eight different QDMs.

\begin{figure}[htbp]
    \begin{center}
        \includegraphics[width=0.8\columnwidth]{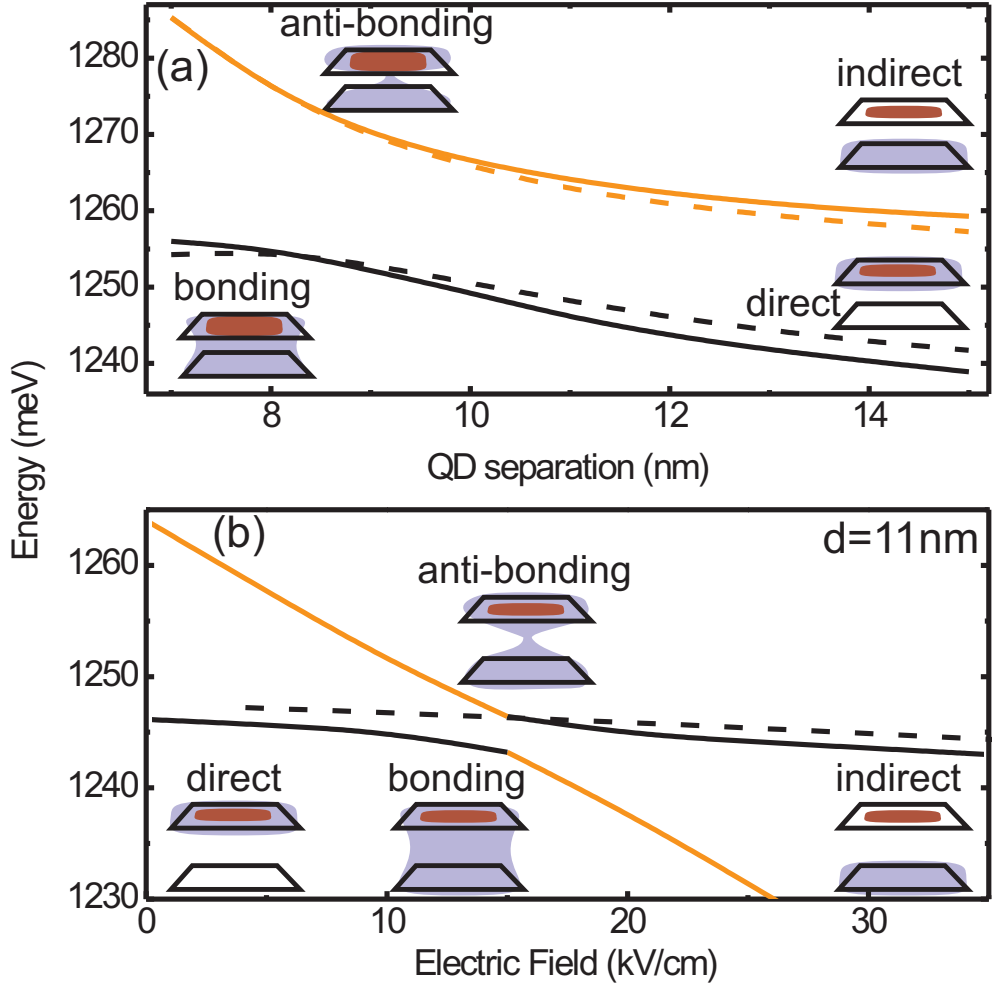}
   \end{center}
  \caption{\label{fig3}(Color online) (a) Exciton energy as a function of dot separation ($d$) for identical dots with the structural parameters described in the text. The direct (black) and indirect (gray) branches are non-degenerate, representing states with the hole in the upper (solid) and lower (dashed) QD. (b) Stark shift for excitonic states with the hole localized in the upper QD for $d=11\mathrm{nm}$. As discussed in the text the ground state undergoes a transition from direct to indirect character at $F\sim 15\mathrm{kV/cm}$ and shows an anti-crossing with the initially indirect exciton. The dashed line is the calculated QCSE of a single dot, for comparison.}
\end{figure}

In order to understand these observations we performed calculations of the single exciton states and their DC Stark shift for our QDMs based on structural information obtained from our TEM-microscopy and previous studies of the composition profile of similar Ga$_{0.5}$In$_{0.5}$As dots.\cite{finley04,Liu} A one band effective mass Hamiltonian was used to calculate the single-particle states including a full treatment of strain and piezoelectric effects. The Coulomb interaction between the electron and hole was treated self-consistently to obtain the excitonic transition energies.\cite{Perdew}
The calculated energies of the four lowest lying excitonic states of such idealized QDMs are presented in Fig. \ref{fig3}(a) as a function of the dot-dot separation ($d$), defined as the separation between the wetting layers of the two dots.

For $d>9\mathrm{nm}$ the tunneling coupling between the two dots is weak and the four exciton states can be readily discussed in terms of a simplified single particle picture.  Within this picture, two species of excitons can be distinguished: {\it direct} excitons, in which the electron and hole are localized in the same dot, and {\it indirect} excitons, in which the electron and hole are in different dots. The direct (indirect) excitons form the lower (upper) pair of states shown as black (gray) lines in Fig. \ref{fig3}(a). The energy separation is due to the attractive Coulomb interaction ($\sim 20 \mathrm{meV}$) which is present only for direct excitons with large electron-hole overlap. The direct and indirect branches themselves are non-degenerate, split by $\sim 3\mathrm{meV}$ due to the long-range influence of the strain field of the lower QD on the bandstructure of the upper.  Within our simplified picture, the two states in each branch correspond to hole localization in the upper or lower dot respectively as depicted schematically by the icons in Fig. \ref{fig3}(a).  For decreasing $d$ the \emph{electron} component of the exciton wavefunction hybridizes into bonding and anti-bonding orbitals whilst the hole remains essentially uncoupled due to the much larger effective mass. For $d\lesssim9\mathrm{nm}$, the electron tunnel coupling energy becomes large ($\gtrsim20\mathrm{meV}$) and, using simpler models, a strong red-shift of the lowest lying exciton state would be expected as reported previously in ref. \cite{Bay}. In strong contrast, the calculated energy of the lowest exciton level in Fig. \ref{fig3}(a) actually exhibits a \emph{blue}-shift as $d$ is decreased.  This effect arises from the combined influence of the reduced electron-hole Coulomb interaction (the electron component of the wavefunction becomes delocalized over both dots whilst the hole hybridization is weak) and the widening of the bandgap due to mutual strain field penetration from one dot into the other. We note that the results of our calculations presented in Fig. \ref{fig3}(a) are in very good agreement with recent pseudopotential calculations that go beyond our one-band model, whilst emphasizing clearly the importance of both strain and Coulomb effects to treat the exciton spectrum of self-assembled QDMs.\cite{Bester} In the limit of weak tunneling coupling ($d>9$nm) higher order effects are negligible and our simulations, presented below, quantitatively describe the field dependence of the exciton transition energies.

For our calculations we focused on similar QDMs to those investigated experimentally ($d=11\mathrm{nm}$) in order to describe quantitatively the different QCSE of the direct and indirect excitons as the electric field ($F$) is tuned. As discussed above the two exciton species in weakly coupled QDMs differ in their spatial electron-hole configurations. The energy peturbation due to the QCSE is given by $\Delta E_{Stark}= p\cdot F$, where $p=e\cdot s$ is the dipole moment of the exciton and $s$ the vertical separation between the centre of the electron and hole charge distributions. For direct excitons the electron and hole wavefunctions are localized in the same QD and separated by $s_{direct}\lesssim 1\mathrm{nm}$ leading to a small excitonic dipole moment $p_{direct}\sim e\cdot s_{direct}$. The magnitude of $p_{direct}$ is \emph{weakly} dependent on $F$ and is not sensitive to the interdot separation ($d$).  This gives rise to a \emph{weak} parabolic Stark shift comparable to single layer samples.\cite{Fry} In contrast, for excitonic states with indirect character the electron-hole separation is much larger ($s_{indirect}\sim d$) and therefore the dipole ($p_{indirect}=e\cdot d$) is large and independent of $F$. From these simple considerations, we expect relative tuning of direct and indirect excitons due to the \textit{strong linear} shift for indirect and the \textit{weak parabolic} shift for direct excitons.\cite{Shen,Fry} This behavior can be seen in Fig. \ref{fig3} (b) where the splitting of the indirect (orange curve) and direct (black curve) states reduces for $0<F(\mathrm{kV/cm})\lesssim10$ according to $\Delta E_{C}\sim E_{C}^{0} - e\cdot (d-s_{direct})\cdot F$ with $E_{C}^{0}\sim 15\mathrm{meV}$. For $F\sim 15\mathrm{kV/cm}$ direct and indirect excitons are tuned into resonance and the electron component of the exciton wavefunction becomes delocalized over both dots, forming symmetric and anti-symmetric molecular states. These two hybridized states are split by the tunneling coupling energy of $\Delta E\sim 5\mathrm{meV}$ for $d=11\mathrm{nm}$ giving rise to the experimentally observed anti-crossing. For $F>15\mathrm{kV/cm}$ the Stark energy of the indirect excitons exceeds the Coulomb interaction of the direct exciton. Therefore, the nature of the QDM ground state changes from direct to indirect as depicted schematically in Fig. \ref{fig3} (b). Since the electron-hole overlap for indirect excitons is negligible, only direct excitons are optically active. At the anti-crossing these two species cannot be distinguished due to electron coupling which leads to \emph{finite} oscillator strength of both branches. For $F>15\mathrm{kV/cm}$ the ground state lo ses oscillator strength, due to its indirect character, as observed experimentally via the field induced quenching of state \textbf{A}. 

We adapted the structural parameters of the model QDMs to obtain the best fit to the experimental data by slightly adjusting the relative size of the dots in the upper and lower layers and the dot-dot separation. The lines of Fig. \ref{fig2}(a) compare our calculations with the experimentally measured peak positions of \A and \B as a function of $F$.  For the fitting, the electric field at which the anti-crossing occurs was found to be most sensitive to the relative vertical height of the lower and upper dots ($h_{l}$ and $h_{u}$) whilst the anti-crossing energy was mainly determined by the dot-dot separation $d$.  The best fit shown in Fig. \ref{fig2}(a) was obtained $h_{l}=3.5\mathrm{nm}$, $h_{u}=4\mathrm{nm}$ and $d=12\mathrm{nm}$ and provides good quantitative agreement with the experimental data.

\begin{figure}[htbp]
    \begin{center}
        \includegraphics[width=0.8\columnwidth]{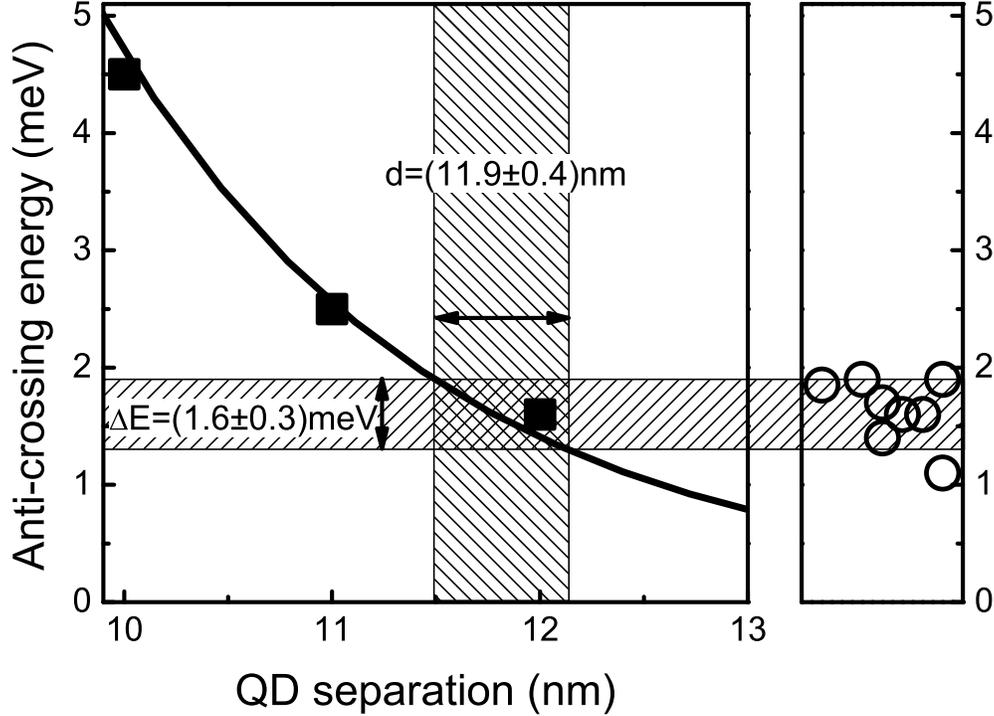}
    \end{center}
    \caption{\label{fig4}Calculated anti-crossing energy as a function of the dot separation (solid line).  The full line shows a fit to the calculated points, describing the exponential nature of the dot-dot coupling strength.  The horizontal shaded bar indicates the standard deviation of the experimental values. The vertical bar represents the dot separation derived form the experimental data and the calculated values of $\Delta E$.}
\end{figure}

We investigated the dependence of the tunnel coupling on $d$ by calculating the anti-crossing energy splitting $\Delta E$ for the model QDMs used to fit the experimental data of Fig. \ref{fig2}.  The results of these calculations are presented in Fig. \ref{fig4} and show that $\Delta E$ decreases exponentially from $4.5\mathrm{meV}$ to $1.5\mathrm{meV}$ as $d$ increases from $10\mathrm{nm}$ to $12\mathrm{nm}$.  This behavior reflects the exponential decay of the tunneling coupling strength for increasing inter-dot separation. In the right hand panel of Fig. \ref{fig4} we plot the energy splitting measured from eight different QDMs and its mean value with the associated standard deviation, $\Delta E= 1.6\pm 0.3\mathrm{meV}$, as indicated by the horizontal shaded region on Fig. \ref{fig4}.  By comparison with our calculations, this value corresponds to an average dot separation of $d= 11.9 \pm 0.4\mathrm{nm}$, in good agreement with the nominal wetting layer separation of $10\mathrm{nm}$ obtained from our TEM investigations.  These findings provide further support for our identification of the observed anti-crossing as arising from a field driven resonance between exciton states of direct and indirect character.

In conclusion, we have demonstrated controlled electronic coupling of excitonic states in an individual QDM via the observation of a electric field induced anti-crossing of spatially \textit{direct} and \textit{indirect} excitons.  The observed coupling strength is in good quantitative agreement with theoretical modeling that includes a realistic treatment of strain and Coulomb interactions. Our results represent the first direct observation of the controlled generation of a molecular state in a QDM by an externally controlled parameter (gate voltage) and may lay the foundations for scaling exciton qubits in semiconductor quantum dots by manipulation of the coherent coupling between QD states.\cite{Villas}

The authors gratefully acknowledge financial support by Deutsche Forschungsgemeinschaft via SFB631 and A. J. Cullis, M. Migliorato and S. L. Liew of the University of Sheffield, U.K. for TEM microscopy.


\begin{thebibliography}
\bibitem {countrate}  Under the extremely weak excitation conditions employed here the typical count rate is only $\sim0.2$ cps per channel.
\end{thebibliography}
\end{document}